\documentclass[twocolumn,showpacs,aps,prl]{revtex4}
\usepackage{epsfig,amssymb}

\begin{document}

\title{Phase Space Manipulation of Cold Free Radical OH Molecules}
\author{ J.~R. Bochinski, Eric R. Hudson, H.~J. Lewandowski, and Jun Ye}
\affiliation{JILA, National Institute of Standards and Technology and University of Colorado and Department of Physics,
University of Colorado, Boulder, Colorado 80309-0440}
\author{Gerard Meijer}
\affiliation{FOM$-$Institute for Plasma Physics Rijnhuizen, P.O.
Box 1207, NL-3430 BE Nieuwegein} \affiliation{Dept. of Molecular
and Laser Physics, University of Nijmegen, Toernooiveld 1,
NL-6525 ED Nijmegen, The Netherlands}
\affiliation{Fritz-Haber-Institut der Max-Planck-Gesellschaft,
Faradayweg 4-6, D-14195 Berlin, Germany}
\date{\today}

\begin{abstract}
We report bunching, slowing, and acceleration of a supersonically
cooled beam of diatomic hydroxyl radicals (OH). \textit{In
situ} observation of laser-induced fluorescence along the beam
propagation path allows for detailed characterization of longitudinal phase-space manipulation of OH molecules through the Stark effect by
precisely sequenced inhomogeneous electric fields.
\end{abstract}

\maketitle

The advent of laser cooling and Bose-Einstein condensation has
transformed atomic physics \cite{nobel}.  Molecules, with their
richer internal structure, offer many new exciting research
opportunities. Cold molecules can be used for novel collision studies, molecular optics and interferometry,
quantum collective effects, quantum information science,
long-range intermolecular states, precision spectroscopy and
measurements, and investigations of chemistry in the ultracold
regime.  At the present stage of development, cold molecular
samples are beginning to be produced and even confined within
traps \cite{knize1998,doyle1998,meijer2000}. The revolutionary
role played by cold atoms to the field of atomic physics may be
duplicated by cold molecules to the field of physical chemistry
and  physical science overall.

The complex structure of molecular energy levels has so far
precluded a straightforward implementation of laser cooling.
Photo-association \cite{heinzen2000} and Feshbach resonance
association of cold atoms \cite{wieman2002} represent effective
approaches to create cold diatomic molecules, but the techniques
are limited to a small class of laser-cooled atoms. Molecules can
also be cooled by thermalization with cryogenically refrigerated
buffer gas, at a small cost of system complexity \cite{doyle1998}.
Working with molecules generated by supersonic expansion has a
definite advantage: the internal - rotational and vibrational -
degrees of freedom of the molecules are cooled and, primarily,
only the ground state is populated. Furthermore, the
translational temperature of the molecular beam is vastly
reduced in its moving frame. Subsequently, the velocity of the
molecules in the laboratory frame can be manipulated by an
elegant method, which takes advantage of the Stark shift
associated with polar molecules \cite{meijer2000b,meijer2002};
inhomogeneous pulsed electric fields controlled with a precise
timing sequence may be used to selectively bunch, slow, or
accelerate the molecules and accumulate them in an ultralow
temperature phase space distribution.

In this Letter we demonstrate the feasibility of this approach
for the diatomic hydroxyl radical (OH).  The selection of the
hydroxyl radical comes from both its extensive relevance to
astrophysics \cite{zadeh2002} and physical chemistry
\cite{yang2000,nesbitt2001} as well as its particular amenability
to the Stark slowing technique. Recent theoretical studies have
also revealed particularly interesting prospects for study of
cold collisions and controlled interactions between OH molecules
\cite{bohn2003}. These molecules possess permanent electric
dipole moments that bring new opportunities for control over
intermolecular interactions.  When the translational energy
becomes comparable to or less than the intermolecular
dipole-dipole interaction energy, polar molecules easily
influence each other's trajectory and orientation, leading to
complex dynamics.  An external applied field can manipulate the
intermolecular interaction: for example, molecules can be made to
collide (or not) on command, by varying the field strength with
respect to the distance-dependent intermolecular dipole
interactions.  This mechanism represents a profoundly different
type of control over collisions than resonant scattering
controlled by a magnetic field. Additionally, the OH molecule possesses a large magnetic moment and is suitable for magnetic trapping.
It will be particularly interesting to explore variable control of
electric fields to manipulate intermolecular interactions in a magnetic trap.
Furthermore, with the existence of a cold, trapped sample, the
small mass of OH radicals should aid a direct laser cooling effort
through vibrational transitions around 2.8 $\mu$m, as cycling
through the lowest vibrational transition requires only three
repumping lasers. Additionally, we note practical, limited laser
cooling along relatively radiatively closed electronically
excited pathways is possible for OH.

A roadblock for cold molecular experiments is the need to develop
and implement means to detect, image, and characterize small
samples of cold molecules once they are trapped. This is a
prerequisite for exploring molecular condensates, coherent
atom-molecule conversion, cold molecule collisions, and ultracold
chemistry.  For this purpose, an important feature of using OH
radicals is ease of detection through laser-induced
fluorescence (LIF). Resonance fluorescence has been developed as
a highly sensitive technique for detecting small numbers of
molecules. By counter-propagating a LIF laser beam along the molecular beam path, we can make definitive \textit{in
situ} observations of the molecular population, density, and
velocity distribution at various positions within the decelerator,
leading to a clear picture of longitudinal phase space evolution.

It is notably difficult to adapt the supersonic expansion
technique that accommodates efficient production of cold, ground
state free radicals via discharge or photolysis and still remain
suitable for the Stark manipulation approach, which relies on the
propagation of a near-collimated molecular beam through
inhomogeneous electric fields. We utilize a high-voltage ($\sim$
3kV) discharge to create OH radicals. Figure 1
shows the experimental setup along with the detected OH signals
along the beam path. Xenon (Xe) gas bubbles through a small tank
of distilled water with a backing pressure of ~2.5 atmospheres.
The pulsed valve has a 0.5 mm diameter nozzle and is operated at
a 5 Hz repetition rate, producing a supersonic beam with pulse
widths of $<$ 100 $\mu$s. Two thin stainless steel disks,
electrically isolated from each other and from earth ground, act
as discharging electrodes with electrons flowing against the
expanding supersonic jet through the central 4 mm opening.
Experimentally we find operating the discharge voltage in a
pulsed mode with a $\sim$2 $\mu$s duration significantly reduces undesirable
heating of the molecular beam. The measured mean velocity of the
beam and the associated spread are 370 m/s $\pm$ 60 m/s,
consistent with a near room temperature, Xe-seeded supersonic
expansion, where the operation of the pulsed valve and discharge
has contributed only slight heating to the molecules. Another
benefit of the short discharging pulse is that the
production of OH molecules is well defined both in time ($<$ 2
$\mu$s) and position ($<$ 2 mm). Trace (a) in the lower panel of
Fig. 1 shows a corresponding discharge light signal that defines
the origin for all timing axes in the subsequent data figures
throughout this Letter. State-selective fluorescence detection
determines that greater than 93\% of OH created are in the
rotational and vibrational ground state (v =0, J = 3/2) of X$^2
\Pi_{3/2}$. Within this state, the symmetrical ``f'' component is
a weak-field seeking state with two fine structure levels of which
the $|$m$_{\mbox {j}}|$ =3/2 component offers three times the
Stark shift of the $|$m$_{\mbox {j}}|$ =1/2.

After traversing a 1.5 mm aperture skimmer, which is charged
(-300V) in order to prevent ions generated in the discharge from
reaching the slower, the molecular beam is focused into the Stark
decelerator \cite{meijer2000b} by a short hexapole. The linear accelerator is formed by 69 pairs of 3.2
mm diameter rounded stainless steel rods with a center-to-center
separation of 5.2 mm, allowing a 2 $\times$ 2 mm$^2$ transverse
acceptance area. Pairs of electrodes define individual slowing
stages. Successive stages are separated by a center-to-center
distance of 5.5 mm and are oriented alternately at 90$^{\circ}$
relative to each other in order to provide transverse guiding. At
the end of the decelerator an electric quadrupole trap is located
$\sim$10 mm from the last decelerator stage.

A frequency-doubled pulsed dye laser ($\sim$ 10 ns pulse
duration) is tuned resonant with the A$^2 \Sigma^+$(v=1)$
\leftarrow$ X$^2\Pi_{3/2}$ (v=0) transition at 282 nm for
fluorescence detection at 313 nm A$^2 \Sigma^+$(v=1)$\rightarrow$
X$^2\Pi_{3/2}$ (v=1) (lifetime~$\tau$=750 ns). The laser beam is
introduced through the 2 mm diameter apertures centered on the
end caps of the quadrupole trap into the decelerator region. The
laser beam counter-propagates down the entire slower through the
skimmer and into the valve region. By design, the vacuum chamber has large solid-angle optical access for
fluorescence collections at multiple positions along
the molecular beam path. The sensitivity of the fluorescence
measurement is enhanced by the favorable 72\% branching ratio
into the detection decay channel. To suppress scattered laser light, we gate the photomultiplier tube
during the excitation laser pulse and collect fluorescence
signals only during the subsequent microsecond scale decay
lifetime. The detected fluorescence signals are converted to
the corresponding molecular numbers using carefully calibrated
parameters such as laser intensity, photon detection efficiency,
decay branching ratio, etc. Along with the present LIF approach,
we are developing a phase-contrast, cavity-enhanced
detection technique with an aim toward study of
cold molecule dynamics in a non-destructive manner.

Following conventions used in linear accelerator physics (for a
more detailed discussion, see Ref. \cite{meijer2000b,meijer2002}), we define
a local spatial coordinate for a molecule located between
two successive stages $\phi = \frac{z}{2L}2\pi$, with \textit{z}=0
defined at the center of the two stages when
the voltages are switched. A molecule that is accelerated
or decelerated has a different equilibrium position $\phi_0$ than
that of a constant speed molecule, which has $\phi_0$ = 0$^{\circ}$. The
corresponding depth of the travelling potential well decreases as
the magnitude of $\phi_0$ increases. With
$\Delta\phi=\phi-\phi_0$  denoting spatial spread and
$\Delta\nu=\nu-\nu_0$ velocity spread around the synchronous
molecules located at the equilibrium position of the moving trap,
the following equations describe the overall phase space
evolution:$\frac{d^2\Delta\phi}{dt^2}+\frac{\pi
W}{mL^2}[\mbox{sin}(\phi_0+\Delta\phi)-\mbox{sin}(\phi_0)]=0$, $\Delta\nu=\frac{L}{\pi}\frac{d\Delta \phi}{dt}$. Here \textit{m} is the molecular mass and \textit{W} denotes the
maximal energy loss (or gain) per stage.

To demonstrate the powerful capability of \textit{in situ}
detection of molecules within the decelerator for the study
of phase space manipulation, we select five representative regions
for LIF signal collection as shown in Fig. 1: (b) before the
skimmer, (c) between the hexapole and the first decelerator
stage, (d) between the 23rd and 24th stages, (e) between 37th and
38th, and (f) between 50th and 51st. Detection at regions (b) and
(c), with the corresponding signals shown in the lower panel of
Fig. 1, allows for optimization of OH production from the valve,
discharge, and hexapole focusing. Experimentally we optimize the
magnitude and duration of the applied hexapole voltage for a
selected velocity class in order to provide transverse
phase-space matching into the decelerator. Applying static
voltage to the decelerator electrodes transversely guides the
molecular beam. The LIF signals detected at (d), (e) and (f), such
as those shown in Fig. 1, are obtained with $\pm$12.5 kV  guiding
voltages applied to all stages, resulting in a $>$10$\times$ signal
enhancement. In all traces shown, the electric fields are
switched off prior to molecular detection.

\begin{figure}
\leavevmode \epsfxsize=3.0in \epsffile{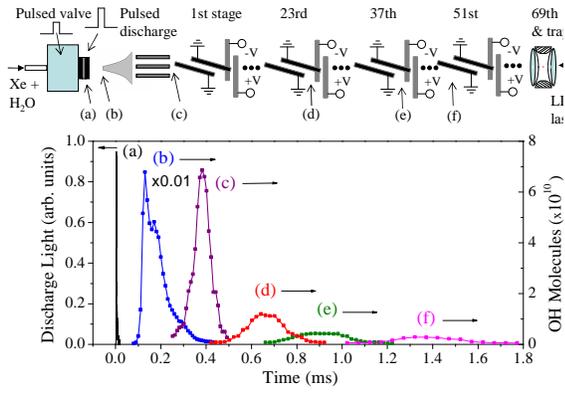}
\caption{\label{fig1}Upper panel: schematic of the Stark
decelerator for free radical OH.  Lower panel: (a) discharging
light at nozzle; LIF signal, (b) before skimmer, (c) after
hexapole, (d) after 23 stages, (e) after 37 stages, (f) before 51
stages. }
\end{figure}

Given good spatial resolution in signal collection, the LIF
signal, presented with respect to the time-of-flight through the
selected detection region, reveals the longitudinal spatial and
velocity distributions, which are input parameters for our model
of the Stark manipulation process.  We image a 2 mm-long beam
section located between two successive pairs of electrodes onto
the PMT, with a magnification of 2.7 to aid spatial filtering.
Figure 2 shows time-of-flight LIF signals observed after 14
stages of inhomogeneous electric fields switched with precise
timing, followed by 9 additional stages of static transverse
guiding. Fig. 2(a) depicts the transversely guided molecular
packet. With the electric fields having the same magnitude and
configuration as in the static guiding case except that the
voltages are now switched at a frequency matching $\nu$/2L,
bunching of molecules is demonstrated in Fig. 2(b). Here $\nu$ is
the mean velocity of the molecular packet (375 m/s)
and \textit{L} is the distance between the successive,
orthogonally oriented stages. The switching effectively creates
travelling potential wells in which molecules are trapped,
producing the central bunched peak marked by the arrow in the
figure. Other peaks and valleys visible in trace (b) are created
by phase-space open trajectories of untrapped molecules. By chirping the
switching frequency, the molecules' speed can be increased to 421
m/s [Fig. 2(c)] or decreased to 323 m/s [Fig. 2(d)], depending on
the sign of the frequency chirp; this results in the central
molecular peak arriving earlier (when accelerated) or later (when
slowed) in time, relative to the bunched peak. Clearly, the
\textit{in situ} detection capability has allowed us to simultaneously monitor,
stage by stage, the trapped and untrapped molecular dynamics, showing temporal transportation of the untrapped
molecules as theoretically investigated in \cite{dong2003}. Our
theoretical simulations (see discussions below) account for these
dynamics in a detailed manner.

\begin{figure}
\leavevmode \epsfxsize=3.0in \epsffile{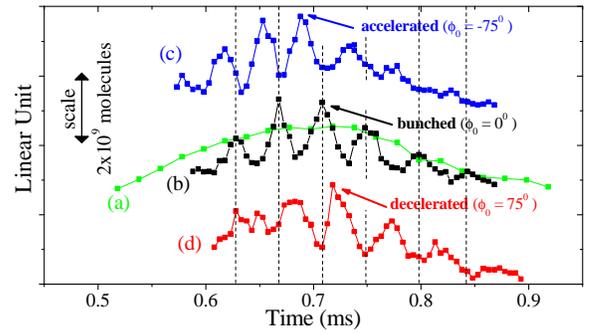}
\caption{\label{fig2}(a) Transverse guiding, (b) bunching, (c)
deceleration, and (d) acceleration of OH after 14 stages of Stark
slowing followed by 9 additional stages of transverse guiding.
For the sake of clarity, the deceleration and acceleration data
are shifted vertically. The peaks marked by arrows represent
molecules trapped in the moving potential wells.}
\end{figure}

Theoretical simulations are derived from detailed understanding
of the Stark slower and OH molecules.  From the measured
geometrical and electrical properties (sizes, distances, and
applied voltages), an electric field map is constructed, which
when coupled with understanding of the OH energy structure
(including both $|$m$_{\mbox{j}}|= 3/2$ and 1/2 levels, with a
mixing ratio of 3:1 resulting from the initial hexapole
selection), gives complete information on the forces experienced
by OH molecules within the decelerator region.  The initial 1-D
phase space distribution is inferred from the static guiding
data, then propagated through the Stark manipulation process
resulting in a phase space distribution similar to those of Fig.
3(c) and (d). This distribution is integrated to produce a
time-of-flight spectra that is compared directly to the
experimental data as shown in Fig. 3(a) and (b).

The LIF experimental configuration is particularly suitable to
visualize effects described by the equations given above.
Molecular packets with known initial velocities can be
decelerated using a pre-determined number of stages, then allowed
to free fly through a certain distance under only transverse
guidance. Subsequent LIF detections reveal the final phase space
distribution, which is related to the decelerated state via a
free flight convolution matrix. A more direct observation of the phase
space manipulation by the slowing stages is achieved by collecting LIF
signals immediately following the deceleration process. Fig. 3(a)
and (b) show two cases where the LIF detections are made after 22
and 36 stages of deceleration at $\phi_0 = 75^{\circ}$, followed
by one stage of free flight.  Corresponding theoretical
simulations are shown as solid lines for both panels. The simulations confirm that the largest peak at 1.02 ms in Fig. 3(b) is associated with the $|$m$_{\mbox {j}}|$ =3/2 state and the second peak at 1.05 ms for $|$m$_{\mbox {j}}|$ =1/2.  The
simulation results are obtained from the phase space distribution
contours shown in the corresponding panels (c) and (d) of Fig. 3,
which also includes a color-coded scale bar for normalized
molecular numbers. As shown in the inset of Fig. 3(c), the
initial phase space distribution ($\Delta v$: 250 - 500 m/s;
$\Delta z$: 0.09 - 0.17 m) at the input of the decelerator is a
result of molecular free flight from the nozzle. Simulations are
carried out with no free parameters except a global scaling
factor. The thickness of the simulation curves signifies the
measurement uncertainty of the geometrical dimensions of the
entire decelerator ($\pm$1 mm over 40 cm), without taking into
account other real-time fluctuations in switch timing, laser
intensity, and molecular numbers. Nevertheless the agreement
between simulations and experimental data is excellent.

In the first case of 22 stages of deceleration, there are five distinct
spatially separated distributions in Fig. 3(c), labelled
respectively as 1, 2, 3, 4, and 5. These distributions fly
through the detection zone indicated by vertical blue lines, leading to the observed LIF
signals shown in Fig. 3(a). It is clear from Fig. 3(c) that
packet 1 represents the trapped molecules that have migrated from
the original phase space distribution. The projected velocity
distribution associated with 1 is also shown near the right
vertical axis, indicating a 7 m/s velocity spread. When the deceleration process is
carried to the 36th stage, the mean velocity of the molecular
packet is further reduced, along with additional rotation of the
phase space distribution, as shown in Fig. 3(d), leading to a
narrower velocity spread of 4.4 m/s. Part of the reduction of the velocity width is attributed to the
departure of untrapped molecules from the packet, as shown in both
experimental data and simulations.

\begin{figure}
\leavevmode \epsfxsize=3.0in \epsffile{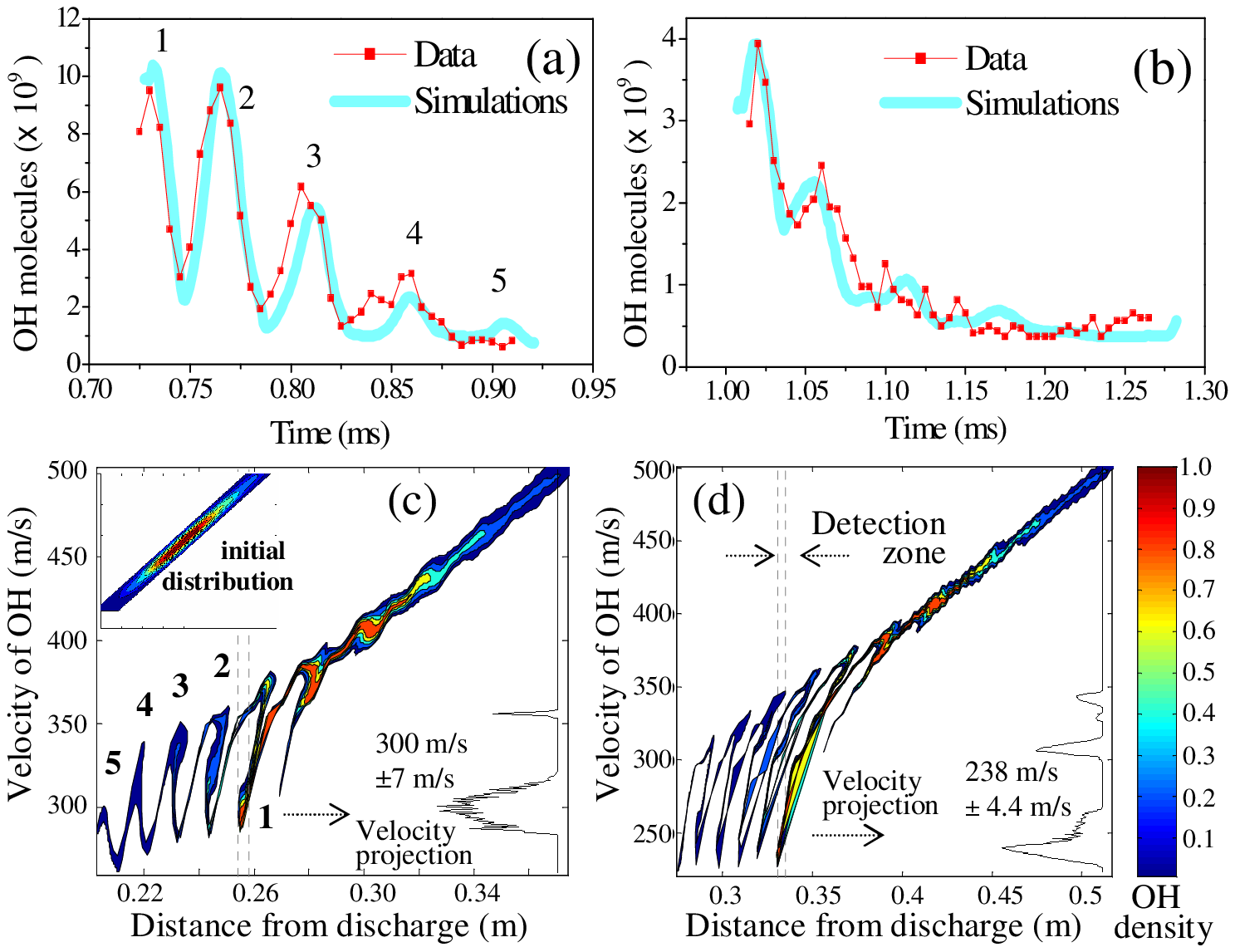}
\caption{\label{fig3}Measured time-of-flight distribution at (a)
23 stages and (b) 37 stages (22 and 36 stages of deceleration,
respectively, followed by guidance for 1 stage). Also shown are
instantaneous pictures of phase space distributions of OH after
(c) 22 and (d) 36 stages of deceleration. In both (a) and (b),
data points are red squares and theory thick blue lines.}
\end{figure}

Finally, as shown in Fig. 4, when the deceleration process is continued to the 50th stage, the
slow molecular packet (red circles) ($\phi_0 = 75^{\circ}$) has now
been completely decelerated out of the transversely
guided molecular distribution (green diamonds). The phase-stable molecular packet has been decelerated to 140 m/s with a velocity spread of 2.7 m/s, corresponding to a longitudinal temperature of 15 mK. Simulations again confirm the position and width of the measured packet, which is now solely attributed to the $|$m$_{\mbox {j}}|$ = 3/2 state. The bunched peak (black squares) ($\phi_0 = 0^{\circ}$) appears 0.3 ms earlier than the slowed packet in the measured time-of-flight distribution.

\begin{figure}
\leavevmode \epsfxsize=3.0in \epsffile{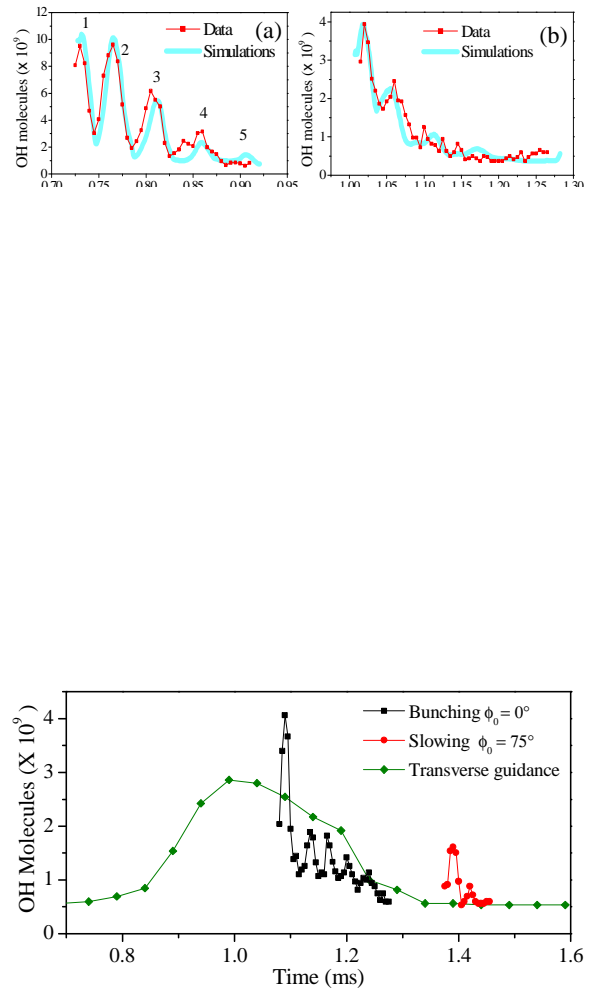}
\caption{\label{fig4}OH molecules after 50 stages of transverse
guidance (green diamonds), bunching (black squares), and
deceleration (red circles).}
\end{figure}

In conclusion, we have presented clear evidence for the first
time of longitudinal phase space manipulations of free radical molecules,
demonstrating bunching, slowing, and acceleration of a ground
rovibrational state of OH. The experimental approach presented here has led to an excellent understanding of the
associated physical processes and will have a major impact on
future cold molecule work where \textit{in situ} detections can be
critical for observations of cooling and trapping dynamics.

We thank J. Bohn, E. Cornell, and D. Nesbitt for
useful discussions. We are grateful to technical helps from A.
Pattee, M. Silva, T. van Leeuwen, and H. Bethlem. This research is supported
by NSF, NIST, and the Keck Foundation.  J. R. B. and H. J. L.
acknowledge support from NRC. J. Ye's email address is
ye@jila.colorado.edu.

\end{document}